\documentclass[14pt,prb,twocolumn]{revtex4}
\usepackage{amsmath,amssymb}
\newcommand{\NN}{\nonumber\\}
\newcommand{\simD}{\stackrel{\text{\tiny D}}{\sim}}
\newcommand{\U}[1]{\mathrm{#1}}
\newcommand{\sub}[1]{_{\mbox{\scriptsize#1}}}
\newcommand{\vct}[1]{\boldsymbol#1}
\newcommand{\ph}[1]{\tilde{#1}}
\newcommand{\fracpd}[2]{\frac{\partial #1}{\partial #2}}
\begin{document}
\title{The vacuum impedance and unit systems}
\author{Masao Kitano}
\email{kitano@kuee.kyoyo-u.ac.jp}
\affiliation{Department of Electronic Science and Engineering,
Kyoto University, Kyoto 615-8510, Japan}
\affiliation{CREST, Japan Science and Technology Agency, Tokyo 103-0028, Japan}
\date{\today}

\begin{abstract}
In electromagnetism, the vacuum impedance $Z_0$ is a universal constant,
which is as important as the velocity of light $c$ in vacuum.
Unfortunately, however, its significance 
does not seem
to be appreciated so well
and sometimes the presence itself is ignored.
It is partly because in the Gaussian system of units, 
which has widely been used
for long time, $Z_0$ is a dimensionless constant and of
unit magnitude.
In this paper, 
we clarify that $Z_0$ is a fundamental parameter
in electromagnetism and plays major roles in the following scenes:
reorganizing the structure of the electromagnetic formula
in reference to the relativity;
renormalizing the quantities toward natural unit systems starting
from the SI unit system;
and defining the magnitudes of electromagnetic units.
\end{abstract}

\keywords{
vacuum impedance, Maxwell's equation,
SI unit system, natural unit system, Gaussian unit system
}

\maketitle

\section{Introduction}
The notion of vacuum impedance was introduced
in late 1930's
by Schelkunoff \cite{schelkunoff} in the study of
wave propagation.
It is defined as the amplitude ratio of
the electric and magnetic fields of plane waves in vacuum,
$Z_0=E/H$, which has the dimension of electrical
resistance.
It is also called
the characteristic impedance of vacuum or
the wave resistance of vacuum.
Due to the historical reasons,
it has been recognized as a special parameter for engineers
rather than a universal physical constant.
Compared with the famous formula for the
velocity of light in terms of the 
vacuum permittivity $\varepsilon_0$ and
the vacuum permeability $\mu_0$,
\begin{align}
  c=\frac{1}{\sqrt{\mu_0\varepsilon_0}},
\label{eq10}
\end{align}
the expression for the vacuum impedance
\begin{align}
  Z_0=\sqrt{\frac{\mu_0}{\varepsilon_0}}
\label{eq20}
\end{align}
is used far less often.
It is obvious when you look up index pages of 
textbooks on electromagnetism.
A possible reason is perhaps that the Gaussian system of units,
in which $Z_0$ is a dimensionless constant and of
unit magnitude, has been
used for long time.


In this paper, we reexamine the structure of electromagnetism
in view of the SI (The International System of Units) system
and find that $Z_0$ plays very important roles as a universal
constant.
In this process we also find that 
a wide-spread belief that the Gaussian system of units
is akin to natural unit systems and suitable for
theoretical studies is not correct.

\section{Relativistic pairs of variables}
In relativity,
the space variable $\vct{x}$ and the time variable $t$ are
combined to form a 4-vector
$(ct, \vct{x})$.
The constant $c$, which has the dimension of velocity,
serves as a factor matching the dimensions of time and
space.
When we introduce a normalized variable $\tau\equiv ct$,
the 4-vector is simplified as
$(\tau, \vct{x})$.

In this form, space and time are represented 
by the same dimension.
This can be done offhandedly by setting $c=1$.
It should be noted, however, this procedure is irreversible and
the dimension for time is lost.
There $c$ becomes dimensionless.
It is better to introduce normalized
quantities such as $\tau$ when we compare
the different systems of units.

When the ratio of two quantities $X$ and $Y$ is 
dimensionless (just a pure number),
we write
$X\simD Y$ and read $X$ and $Y$ are dimensionally equivalent.
For example, we have $ct\simD\vct{x}$.
If a quantity $X$ can be measured in
a unit $\U{u}$,
we can write $X\simD\U{u}$.\cite{notation}
For example, for
$l=2.5\,\U{m}$ we have $l \simD \U{m}$.

With this notation,
we can repeat the above discussion.
For $ct\simD x$,
instead of recasting $t\simD x$ by
forcibly setting $c=1$,
we introduce a new normalized quantity $\tau=ct$ and have
$\tau\simD x$.
Then velocity $v$ and $c$ is normalized as
$\tilde{v}=v/c$ and $\tilde{c}=c/c=1$, respectively.

From the relativistic point of view,
the scalar potential $\phi$ and the vector potential $\vct{A}$ 
are respectively a part of a unified quantity.
Considering $\phi\simD\U{V}$, $\vct{A}\simD\U{Vs/m}$,
we can introduce a pair of quantities:
\begin{align}
(\phi, c\vct{A})\simD \U{V}
\label{eq200}
.
\end{align}
Similarly, we introduce other pairs:
\begin{align}
& (\vct E, c\vct B)\simD \U{V/m},\NN  
& (\vct H, c\vct D)\simD \U{A/m},\NN  
& (\vct J, c\varrho)\simD \U{A/m^2}
\label{eq400}
.
\end{align}
where $\vct E$, $\vct B$, $\vct H$, $\vct D$,
$\vct J$, and $\varrho$ represent
electric field, magnetic flux density,
magnetic field strength, electric flux density,
and charge density, respectively.
Mathematically those pairs are
anti-symmetric tensors in
the 4-dimensional space as defined
in Appendix.

We have seen that the constant $c$ appears when
we form a relativistic tensor from a pair of
non-relativistic electromagnetic quantities.
In Table \ref{tab100},
such relativistic pairs are listed
according to their tensor orders.
We will find that 
this table is very helpful to overview the structure 
of electromagnetism.
\cite{sommerfeld,deschamps}

\begin{table}
  \centering
\begin{tabular}{ccc}
\underline{F series} & & \underline{S series}
\\[2mm]
$(\phi, c\vct{A})$ & & \\[1mm]
\phantom{$\mathsf{d}$} $\downarrow$ $\mathsf{d}$ 
  & \phantom{AB}$Y_0$ & \\[1mm]
$(\vct E, c\vct{B})$ & $\leftarrow$ $*$ $\rightarrow$ 
  & $(\vct H, c\vct D)$ \\[1mm]
\phantom{$\mathsf{d}$} $\downarrow$ $\mathsf{d}$ 
   & $Z_0$\phantom{AB} & 
\phantom{$\mathsf{d}$} $\downarrow$ $\mathsf{d}$ 
\\[1mm]
0 & &    $(\vct{J}, c\varrho)$ \\[1mm]
  & & 
\phantom{$\mathsf{d}$} $\downarrow$ $\mathsf{d}$ 
\\[1mm]
  & & 0
\end{tabular}
\caption{
Relativistic pairs of quantities are
arranged as a diagram, the rows of which correspond to
the orders of tensors ($n=1,2,3,4$).
In the left column, the quantities related to
the electromagnetic forces (the F series),
and in the right column, the
quantities related to the
electromagnetic sources (the S series) are listed.
The exterior derivative ``$\mathsf{d}$'' connects 
two pairs of quantities.
These differential relations correspond to
the definition of (scalar and vector) potentials,
the Maxwell's four equations, and the charge conservation. 
Hodge's star operator ``$*$'' connects
$(\vct E,c\vct B)$ and $(\vct H, c\vct D)$ pairs.
This corresponds to the constitutive relations for
vacuum.
Here appears the vacuum impedance $Z_0=1/Y_0$.
}
\label{tab100}
\end{table}

\section{Roles of the vacuum impedance}
In this section we show
some examples for which $Z_0$ plays important roles.
\paragraph{Source-field relation}
We know that the scalar potential $\varDelta\phi$
induced by a charge $\varDelta q=\varrho\varDelta v$ is
\begin{align}
\varDelta\phi = \frac{1}{4\pi\varepsilon_0}\frac{\varrho\varDelta v}{r}
,
\label{eq600}
\end{align}
where $r$ is the distance between 
the source and the point of observation
The charge is presented
as a product of 
charge density $\varrho$ and a small volume $\varDelta v$.
Similarly a current moment (current by length)
$\vct{J}\varDelta v$ 
generates 
the vector potential
\begin{align}
\varDelta\vct{A} = \frac{\mu_0}{4\pi}\frac{\vct{J}\varDelta v}{r}
.
\label{eq610}
\end{align}
Considering the relations (\ref{eq200}) and (\ref{eq400}),
these equations can be united as
\begin{align}
  \begin{bmatrix}
\varDelta \phi\\
\varDelta(c\vct{A}) 
  \end{bmatrix}
=
\frac{Z_0}{4\pi r}
\begin{bmatrix}
  c\varrho\\
 \vct J
\end{bmatrix}
\varDelta v
\label{eq700}
.
\end{align}
We see that the vacuum impedance $Z_0$ plays the role to
relate the source $(\vct J, c\varrho)\varDelta v$ and the resultant
fields $\varDelta(\phi, c\vct{A})$ in
a unified manner.

\paragraph{Constitutive relation}
A similar role can be seen for the constitutive
relations for vacuum;
$\vct{D}=\varepsilon_0\vct{E}$ and
$\vct{H}=\mu_0^{-1}\vct{B}$ can be combined as
\begin{align}
  \begin{bmatrix}
    \vct E \\
    c\vct B
  \end{bmatrix}
=
Z_0
\begin{bmatrix}
  c\vct D\\
  \vct H
\end{bmatrix}
.
\label{eq800}
\end{align}
More precisely, with Hodge's star operator ``$*$''
(see Appendix),\cite{frankel, hehl}
it can be written as 
\begin{align}
(\vct E, c\vct B)=*Z_0(\vct H, c\vct D)
.
\label{eq810}
\end{align}
It should be noted that
the electric relation 
and the magnetic relation 
are united
under the sole parameter $Z_0$.

\paragraph{Plane wave}
For linearly polarlized plane waves in vacuum,
a simple relation $E=cB$ holds.
If we introduce $H$ $(=\mu_0^{-1}B)$ instead of $B$,
we have $E=Z_0 H$.
The latter relation was introduced by Schelkunoff \cite{schelkunoff}
in 1938.
The reason why $H$ is used instead of $B$ is as follows.
The boundary conditions for magnetic fields 
at the interface of media 1 and 2 are
$H\sub{1t}=H\sub{2t}$ (tangential) and $B\sub{1n}=B\sub{2n}$ (normal).
For the case of normal incidence, which is most
important practically, the latter condition becomes
trivial and cannot be used.
Therefore $H$ is used more conveniently.
The mixed use of the quantities ($E$ and $H$) of the F and S series invite
$Z_0$ unintentionally.

\paragraph{Magnetic monopole}
Let us compare the force between charges $q$
and that between the magnetic monopoles $g$ ($\simD \U{Vs}=\U{Wb}$).
If these forces are the same for equal distances $r$, i.e., 
$q^2/(4\pi\varepsilon_0 r^2)=g^2/(4\pi\mu_0 r^2)$,
we have the relation $g= Z_0 q$.
With this relation in mind,
the Dirac monopole $g_0$,
whose quantization condition is
$g_0e=h$, 
can be beautifully expressed in terms of
the elementary charge $e$ as
\begin{align}
  g_0=\frac{h}{e}=\frac{h}{Z_0 e^2}(Z_0 e)=(2\alpha)^{-1}Z_0 e
\end{align}
where $h=2\pi\hbar$ is Planck's constant.
The dimensionless parameter
$\alpha=Z_0 e^2/2h = e^2/4\pi\varepsilon_0\hbar c\sim 1/137$ is
called the fine structure constant,
whose value is independent of
unit systems and characterize the strength of
the electromagnetic interaction.
The use of $Z_0$ helps to keep SI-formulae in simple forms.

\paragraph{The F series versus the S series}
Impedance (resistance) is a physical quantity
by which voltage and current are related.\cite{dimension}
In the SI system, the unit for voltage is $\U{V}(=\U{J/C})$ (volt)
and the unit for current is $\U{A}(=\U{C/s})$ (ampere).
We should note that the latter is proportional to
and the former is inversely proportional to the unit of charge,
$\U{C}$ (coulomb).
We also note in Table \ref{tab100} that the units for 
quantities in the F series
are proportional to the volt and those in the S series
are proportional to the ampere.
After all, the vacuum impedance $Z_0$ plays the role to
connect the quantities in the F and S series.
In the above cases we have found that
the use of $Z_0$ (together with $c$) instead of $\varepsilon_0$
or $\mu_0$ simplifies equations.

\begin{table*}
\centering
\begin{tabular}{ll}
\hline
$(\phi, \ph{\vct{A}})$ & $\U{V}$\\
$(\vct E, \ph{\vct{B}})$ & $\U{V/m}$\\
$(\vct{H}, \ph{\vct D})$ & $\U{A/m}$\\
$(\vct{J}, \ph\varrho)$ & $\U{A/m^2}$\\
\hline
\multicolumn{2}{c}{(a) $c$-normalization}
  \end{tabular}
\hspace{2em}
  \begin{tabular}{ll}
\hline
$(\phi, \ph{\vct{A}})$ & $\U{V}$\\
$(\vct E, \ph{\vct{B}})$ & $\U{V/m}$\\
$(\vct{H}^*, \ph{\vct D}^*)$ & $\U{V/m}$\\
$(\vct{J}^*, \ph\varrho^*)$ & $\U{V/m^2}$\\
\hline
\multicolumn{2}{c}{(b) $(c, Z_0)$-normalization}
  \end{tabular}
\hspace{2em}
  \begin{tabular}{ll}
\hline
$(\hat\phi, \acute{\vct A})$ & $\U{\sqrt{N}}$\\
$(\hat{\vct E}, \acute{\vct B})$ & $\U{\sqrt{N}/m}$\\
$(\acute{\vct H}, \hat{\vct D})$ & $\U{\sqrt{N}/m}$\\
$(c^{-1}\hat{\vct J}, \hat\varrho)$ & $\U{\sqrt{N}/m^2}$\\
\hline
\multicolumn{2}{c}{(c) Gaussian}
  \end{tabular}
\hspace{2em}
  \begin{tabular}{ll}
\hline
$(\hat\phi, \acute{\vct A})$ & $\U{\sqrt{N}}$\\
$(\hat{\vct E}, \acute{\vct B})$ & $\U{\sqrt{N}/m}$\\
$(\acute{\vct H}, \hat{\vct D})$ & $\U{\sqrt{N}/m}$\\
$(\acute{\vct J}, \hat\varrho)$ & $\U{\sqrt{N}/m^2}$\\
\hline
\multicolumn{2}{c}{(d) modified Gaussian}
  \end{tabular}
\caption{
Pairs of quantities in electromagnetism and their units (dimension).
Quantities $X$ normalized with $c$ and $Z_0$ are marked as
$\ph{X}=cX$ and $X^*=Z_0X$, respectively.
As seen in (a) and (b), 
the variety of units is reduced
owing to the normalization.
Gaussian unit system (c) and the modified Gaussian unit 
system (d) are presented with normalized variables:
$\hat{S}=S/\sqrt{\varepsilon_0}$,
$\hat{F}=F\sqrt{\varepsilon_0}$,
$\acute{S}=S/\sqrt{\mu_0}$,
$\acute{F}=F\sqrt{\mu_0}$,
where $S$ ($F$) represents a quantity in the S (F) series.
We notice an irregularity in the fourth row of the Gaussian system.
}
\label{tab200}
\end{table*}

\section{The magnitude of the unit of resistance}
Here we consider a hypothetical situation where 
we are allowed to redefine 
the magnitudes of units in electromagnetism.

The product of the unit of voltage and that of current
should yield the unit of power, 
$1\,\U{V}\times 1\,\U{A}=1\,\U{W}=1\,\U{J/s}$,
which is a fixed quantity determined mechanically,
or outside of electromagnetism.
Thus, a new volt $\U{V'}$ and a new ampere $\U{A'}$ must
be defined so as to satisfy
\begin{align}
  \U{A'}=k\U{A},\quad \U{V'}=k^{-1}\U{V}
,
\label{eq900}
\end{align}
in terms of the currently used $\U{V}$ and $\U{A}$,
where $k$ ($\neq 0$) is a scaling factor.
Accordingly a new ohm $\U{\Omega}'$ must be redefined as 
\begin{align}
  \U{\Omega'}=k^{-2}\U{\Omega}
.
\label{eq950}
\end{align}

We denote the numerical value as
$\{A\}\equiv A/\U{u}$, when
we measure a physical quantity $A$
with a unit $\U{u}$.
For example, for $l=1.3\,\U{m}$
we write $\{l\}=l/\U{m}=1.3$.
We can have another numerical value 
$\{A\}'\equiv A/\U{u'}$, when we measure
the same quantity $A$ in a different unit $\U{u'}$.
Now we have the relation
\begin{align}
  A = \{A\}\U{u}=\{A\}'\U{u'}.
\label{eq1000}
\end{align}
It should be stressed that
the physical quantity $A$ itself is independent
of the choice of units.
What depends on the choice is the numerical value
$\{A\}$.

In the SI system,
from the definition of $c$ and $\mu_0$,
the vacuum impedance is represented as
$Z_0=\sqrt{\mu_0/\varepsilon_0}=c\mu_0=
(299\,792\,458\,\U{m/s})\times(4\pi\times10^{-7}\,\U{H/m})
=(119.916\,983\,2\times\pi)\,\U{\Omega}
\sim 377\,\U{\Omega}$.
In our new system, with (\ref{eq950}) we have
\begin{align}
  Z_0 = \{Z_0\}\U{\Omega}=k^2\{Z_0\}\U{\Omega'}=\{Z_0\}'\U{\Omega'}
\label{eq1050}
.
\end{align}
The numerical value must be changed
from  $\{Z_0\}=377$ to $\{Z_0\}'=377k^2$.
For example, we could choose a new ohm $\U{\Omega'}$ so that
$Z_0=1\,\U{\Omega'}$ is satisfied by setting
$k\sim 1/\sqrt{377}$.
Conversely, to fix $\{Z_0\}$ to
a particular number implies the determination of the magnitude of
units ($\U{\Omega}$, $\U{V}$, $\U{A}$, and others) in 
electromagnetism.

Once $k$, or $\{Z_0\}$ is fixed, 
the numerical values for quantities in the F series are multiplied
by $k$ and those in the S series are divided by $k$.
The sole parameter $k$ or $\{Z_0\}$ determines
the numerical relation between the F and S series.

Coulomb's law for charges $q_1$ and $q_2$ can be rewritten as
\begin{align}
  F &= \frac{1}{4\pi\varepsilon_0}\frac{q_1q_2}{r^2}
= q_2 E
= \{q_2\}\U{As}\times\{E\}\U{V/m}
\NN
&=\{q_2\}'\U{A's}\times\{E\}'\U{V'/m}
\label{eq1400}
\end{align}
where 
$E=(4\pi\varepsilon_0)^{-1}(q_1/r^2)$ is the electric field
induced by $q_1$.
We see
\begin{align}
\{q_2\}'=k^{-1}\{q_2\},
\quad
\{E\}'=k\{E\},
.
\label{eq1450}
\end{align}
and find that
the numerical value for the charge $q$ and that for the electric
field $E$ will be changed reciprocally.
We also note $\{\varepsilon_0\}'=k^{-2}\{\varepsilon_0\}$
and $\{\mu_0\}'=k^2\{\mu_0\}$.

In the SI, the ampere is defined in terms of
the force $F$ between the parallel two wires 
carrying the same amplitude of current $I$.
We have $F/l = \mu_0 I^2/(2\pi r)$, where $r$ is the separation and
$l$ is the length of wires.
Substituting $F=2\times10^{-7}\,\U{N}$, 
$r = l = 1\,\U{m}$, 
$I=1\,\U{A}$, we get $\mu_0=4\pi\times10^{-7}\,\U{H/m}$.
Thus the magnitude $\{\mu_0\}$ (or $\{Z_0\}$) are fixed.

We could determine $\{\varepsilon_0\}$ by the
force between charges with the same magnitude.
In Giorgi's unit system (1901), which is a 
predecessor of the MKSA unit system or the SI system,
$k$ was fixed by determining the magnitude of the ohm.
The way of determination $\{Z_0\}$ has 
been and will be changed according to the
development of high precision measurement technique.

\section{Toward natural unit systems}
As shown in Table \ref{tab200} (a),
by introducing a new set of normalized quantities,
$\ph{X}=cX$, derived from SI quantities $X$,
we can reduce the number of fundamental dimensions.
In this case, we only need three; the ampere, the volt, and the meter.

Further, as seen in Table \ref{tab200} (b),
when we introduce a set of normalized quantities,
$X^*=Z_0X$, by multiplying $Z_0$,
only the volt and the meter are required.
By normalizing the quantities with the fundamental
constants, $c$ and $Z_0$,
we have a simplified set of Maxwell's equations:
\begin{align}
&\vct{\nabla}\cdot\ph{\vct D}^* = \ph\varrho^*,
\quad
\vct{\nabla}\times\ph{\vct H}^* = \fracpd{\ph{\vct D}^*}{\tau}
+\vct J^*
\NN
&\vct{\nabla}\cdot\ph{\vct{B}}=0,
\quad
\vct{\nabla}\times\vct E = -\fracpd{\ph{\vct{B}}}{\tau}
.
\label{eq1500}
\end{align}
with $\ph{\vct D}^*=\vct E$ and $\vct{H}^* = \ph{\vct B}$.
Considering $\tau=ct$,
this set of equations resembles to
the Maxwell's equations
in the Gaussian system of units
except for the rationalizing factor $4\pi$
[See Eq.~(\ref{eq1950})].
However there is a significant difference;
the factor $1/c$ is missing in the
current density term.
We will return to this point later.
It should be stressed that
a natural system of units can be reached from
the SI system by normalizations without
detouring via the Gaussian system.

The number of basic units has been reduced from 
four ($\U{m}$, $\U{kg}$, $\U{s}$, $\U{A}$) to
two ($\U{m}$, $\U{V}$) by introducing 
the quantities normalized with $c$ and $Z_0$.
For further reduction toward a natural 
unit system,\cite{trialogue,wilczek}
$\hbar$ and the gravitational constant $G$ can be
used for example.

\section{Gauss and
Heaviside-Lorentz systems of units}

The SI and the cgs (esu, emu, Gaussian) systems differ in three respects.
First,
in the cgs unit systems,
no fundamental dimensions are supplemented
to the three fundamental dimensions for mechanics;
length, mass, and time.
On the other hand in the SI (MKSA) system,
a new fundamental dimension that for electric current
is introduced.
The cgs systems contain three basic units, while
the SI system contains four.

Secondly,
the cgs systems are irrational systems;
the factor $(1/4\pi)$ is erased from Coulomb's
law but the factor $4\pi$ appears in the
source terms of Maxwell's equations instead.
The SI is a rational system, which has the opposite appearance.

Thirdly,
the base mechanical system for the cgs systems is the cgs
(centimeter, gram, and second)
mechanical system.
That for the SI system is the MKS (meter, kilogram, and second) system.

In order to focus all our attention on the first respect,
i.e., the number of basic units,
we will ignore the differences in the last two respects.
{\em From now on, we pretend that all the cgs systems 
(esu, emu, and Gaussian) are
constructed rationally on the MKS mechanical system.}
(Actually the Heaviside-Lorentz system is an MKS version
of the Gaussian system, namely,
a three-unit, rational system based on the MKS system.)

To go from
the SI system to the cgs systems,
we have to reduce the number of basic units by normalization
with a universal constant.

In the cgs electrostatic system of units (esu), 
Coulomb's law is expressed as
\begin{align}
  F
=\frac{1}{4\pi\varepsilon_0}\frac{q_1q_2}{r^2}
=\frac{1}{4\pi}\frac{\hat q_1 \hat q_2}{r^2}
.
\label{eq2150}
\end{align}
Thus, the normalized charge \cite{realcgs}
\begin{align}
  \hat q = \frac{q}{\sqrt{\varepsilon_0}}
\ \simD \frac{\U{C}}{\sqrt{\U{F/m}}}=\sqrt{\U{Jm}}=\U{\sqrt{N}m}
\label{eq2200}
\end{align}
is a quantity expressed by mechanical dimensions only.
The quantities in the S series, each of which is
proportional to the coulomb, $\U{C}$,
can be normalized by division with 
$\sqrt{\varepsilon_0}$.
On the other hand, the quantities in the F series, each of which is
inversely proportional to C, 
can be normalized by multiplication with 
$\sqrt{\varepsilon_0}$.
For example,
$\vct E$,
$\vct D$,
$\vct B$, and
$\vct H$,
are normalized as
\begin{align}
&  \hat{\vct E} = 
    \vct E\sqrt{\varepsilon_0} \ \simD \frac{\sqrt{\U{N}}}{\U{m}},
\quad
   \hat{\vct D} = 
    \frac{\vct D}{\sqrt{\varepsilon_0}}\ \simD \frac{\U{\sqrt{N}}}{\U{m}}
,\NN
&  \hat{\vct B} = 
    \vct B\sqrt{\varepsilon_0} \ \simD \frac{\sqrt{\U{N}}\U{s}}{\U{m^2}},
\quad
   \hat{\vct H} = 
     \frac{\vct H}{\sqrt{\varepsilon_0}}\ \simD \frac{\U{\sqrt{N}}}{\U{s}}
,
\label{eq2300}
\end{align}
respectively.
We have the constitutive relation
\begin{align}
  \hat{\vct D}= \hat{\vct E},
\quad
  \hat{\vct H}= c^2\hat{\vct B},
\end{align}
and the normalized
permittivity 
$\hat\varepsilon_0=1$
and 
permeability
$\hat\mu_0={1}/{c^2}$.
The normalized vacuum impedance is
\begin{align}
  \hat Z_0 = \sqrt{\frac{\hat\mu_0}{\hat\varepsilon_0}}=\frac{1}{c}
\ \simD \frac{\U{s}}{\U{m}}
.
\end{align}

For the cgs electromagnetic system of units (emu),
S-series quantities are multiplied by $\sqrt{\mu_0}$ and
F-series quantities are divided by $\sqrt{\mu_0}$.
With this normalization,
$\mu_0$ is eliminated from the magnetic Coulomb law
or the law of magnetic force between currents.
The fields are normalized as
\begin{align}
&  \acute{\vct B} = 
    \frac{\vct B}{\sqrt{\mu_0}} \ \simD  \frac{\U{\sqrt{N}}}{\U{m}},
\quad
  \acute{\vct H} = 
    \vct H\sqrt{\mu_0} \ \simD \frac{\U{\sqrt{N}}}{\U{m}},
\NN
&  \acute{\vct E} = 
    \frac{\vct E}{\sqrt{\mu_0}} \ \simD \frac{\U{\sqrt{N}}}{\U{s}},
\quad
  \acute{\vct D} = 
    \vct D\sqrt{\mu_0} \ \simD  \frac{\U{\sqrt{N}s}}{\U{m^2}}
.
\label{eq2500}
\end{align}
The constitutive relations are
\begin{align}
\acute{\vct H}=\acute{\vct B},
\quad
\acute{\vct D}=c^{-2}\acute{\vct E}
,
\end{align}
and we have the normalized permeability
$\acute\mu_0=1$
and permittivity 
$\acute\varepsilon_0={1}/{c^2}$.
The normalized vacuum impedance is
\begin{align}
  \acute Z_0 = \sqrt{\frac{\acute\mu_0}{\acute\varepsilon_0}}=c
\ \simD \frac{\U{m}}{\U{s}}
.
\end{align}

The Gaussian system of units is a combination of
the esu and emu systems.
For electrical quantities the esu normalization is used
and 
for magnetic quantities the emu normalization is used.
Namely we use
$\hat{\vct E}$,
$\hat{\vct D}$,
$\acute{\vct B}$,
and $\acute{\vct H}$, all of which
have the dimension $\sqrt{\U{N}}/\U{m}$.
The constitutive relations are simplified as
\begin{align}
\hat{\vct D}=\hat{\vct E},
\quad
\acute{\vct H}=\acute{\vct B},
\label{eq2050}
\end{align}
and we have
$\hat\varepsilon_0=1$ and
$\acute\mu_0=1$.
So far it looks nice because electric and magnetic
quantities are treated symmetrically.
This appearence is the reason why the Gaussian system has been used
so widely.
However, there is an overlooked problem
in the normalization of current density.
It is normalized as
$\hat{\vct{J}}=\vct{J}/\sqrt{\varepsilon_0}$ in the Gaussian system.
The current density is the quantity primarily connected to
magnetic fields and
therefore it should be normalized as
$\acute{\vct{J}}=\vct{J}\sqrt{\mu_0}$ 
as for the emu system.
Because of this miscasting,
we have an irregularity in the fourth row of the column (c)
of Table \ref{tab200}.
The Gaussian normalization happens to make
the pairs of quantities relativistic
with exception of the $(\vct J, \varrho)$ pair.

The relativistic expression for the conservation of charge should be
\begin{align}
  \fracpd{\ph\varrho^*}{(ct)}+\vct{\nabla}\cdot\vct{J}^*=0,\quad
\label{eq1800}
\end{align}
as for the cases of $c$-\ or $(c, Z_0)$-normalization.
In the Gaussian system, however, the non-relativistic
expression
\begin{align}
  \fracpd{\hat\varrho}{t}+\vct{\nabla}\cdot\hat{\vct{J}}=0
\label{eq1900}
\end{align}
is adopted.
As a practical system of units, it is a reasonable 
(and perhaps unique) choice.

This quirk can clearly be seen,
when we compare the Maxwell's equations (\ref{eq1500})
in the natural system of units and
that for the Gaussian system:
\begin{align}
&\vct{\nabla}\cdot\hat{\vct D} = \hat{\varrho},
\quad
\vct{\nabla}\times\acute{\vct H} = \frac{1}{c}\fracpd{\hat{\vct D}}{t}
+\frac{1}{c}\hat{\vct J},
\NN
&\vct{\nabla}\cdot\acute{\vct B}=0,
\quad
\vct{\nabla}\times\hat{\vct E} = -\frac{1}{c}\fracpd{\acute{\vct B}}{t}
.
\label{eq1950}
\end{align}
The factor $1/c$ in the current density term
is a seam introduced when the esu and emu systems are joined
into the Gaussian system.

The common belief that
the Gaussian system is superior to the SI system
because of the similarity to a natural unit system
or because of the compatibility with relativity is almost pointless.
We should remember that the Gaussian unit system
was established in 1870s, when the relativity or
the Lorentz transformation were not known yet.

The modified Gaussian system,\cite{modify} in which 
$\acute{\vct J}$ is adopted and the above seam
is eliminated, has been proposed but is rarely used.
Column (d) of Table \ref{tab200} contains
quantities in the modified Gaussian system.
They differ uniformly by a factor
$\sqrt{\varepsilon_0}$ from 
the $(c, Z_0)$-normalized quantities in Column (b).

\section{Summary and Discussion}
The important expression (\ref{eq10}) for the velocity of light
also holds for the esu and emu systems:
\begin{align}
c=\frac{1}{\sqrt{\mu_0\varepsilon_0}}
=\frac{1}{\sqrt{\hat\mu_0\hat\varepsilon_0}}
=\frac{1}{\sqrt{\acute\mu_0\acute\varepsilon_0}}
,
\label{eq2000}
\end{align}
but not for the Gaussian system;
${1}/{\sqrt{\hat\varepsilon_0\acute\mu_0}}=1\neq c$.

The expression (\ref{eq20}) for the vacuum impedance,
is rewritten as
$\hat{Z}_0=1/c$ and $\acute{Z}_0=c$ 
for the esu and emu systems, respectively.
Maxwell himself worked with the emu system
when he found that light is an electromagnetic
disturbance propagated according to electromagnetic
laws.\cite{maxwell}
For the emu system the dimensions of resistance and
velocity are degenerate.
For the Gaussian system,
the vacuum impedance 
reduces to unity,
$\sqrt{\acute\mu_0/\hat\varepsilon_0} = 1$.

Thus there is no room for the vacuum impedance
in the cgs systems, which contains
only three basic units.
However
when we move to a unit system with four basic units,
the vacuum impedance $Z_0$ should be recognized as a fundamental constant
as important as the velocity of light $c$.
It has been underestimated or ignored for long time.
It is due to the fact that 
the Gaussian system, for which $Z_0$ is
dimensionless and of unit magnitude, has been used for
long time even after the introduction of the MKSA and
the SI systems.

As has been pointed out by Sommerfeld,\cite{sommerfeld}
the Gaussian system tends to veil the significance
of $D$ and $H$ in vacuum.
Sometimes it is told that
in vacuum only $E$ and $B$ have their significance and
$D$ and $H$ lose their meaning. 
This argument is strongly misled
by the use of Gaussian system of units.
Considering the tensorial nature of quantities as in (\ref{eq810}),
the constitutive relations
for the Gaussian system are expressed as
\begin{align}
(\hat{\vct E},\acute{\vct B})=*(\acute{\vct H}, \hat{\vct D})
.
\end{align}
with Hodge's star operator.
This relation represents important geometrical relations
of electromagnetic fields,
which can hardly be suggested by 
the simple vector relations ($\ref{eq2050}$).

Now we have understood that
without the help of Gaussian system,
we can reach natural systems of units
directly from the SI system.
We believe it's time to say goodbye to the Gaussian
system of units.
You won't miss its simpleness if you have the vacuum impedance
$Z_0$ as a key parameter.

\begin{acknowledgments}
We thank K.~Shimoda for helpful discussions.
This work is supported by the 21st Century COE program No.~14213201.

\end{acknowledgments}

\appendix*

\section{tensor notations}
In this appendix, we will explain mathematically
the entities of Table \ref{tab100}\@.
With basis vectors
$\{\vct{e}_0,\vct{e}_1,\vct{e}_2,\vct{e}_3\}$,
a space-time vector can be represented as
\begin{align}
(ct, \vct{x})=(ct)\vct{e}_0+\vct x,
\quad \vct x = \sum_{i=1}^3 x^i \vct{e}_i
.
\end{align}
The basis vectors satisfy
$\vct{e}_{\mu}\cdot\vct{e}_{\nu}=g_{\mu\nu}$ $(\mu,\nu=0,1,2,3)$.
The nonzero elements of $g$ are $g_{00}=-1$, $g_{11}=g_{22}=g_{33}=1$.
We also introduce
the dual basis vectors:
$\{\vct{e}^0,\vct{e}^1,\vct{e}^2,\vct{e}^3\}$,
with $\vct{e}^0=-\vct{e}_0$,
$\vct{e}^i=\vct{e}_i$ $(i=1,2,3)$.

The quantities in electromagnetism are
expressed by antisymmetric tensors of rank $n$
($n$-forms)
in the four-dimensional space.\cite{deschamps,frankel,hehl}
For scalar fields $\alpha$, $\beta$ and
3-dimensional vector fields $\vct X$, $\vct Y$,
$n$-forms $(n=1,2,3)$ are defined as
\begin{align}
(\alpha; \vct X)_1
& =\alpha\vct e^0 + \vct X
,
\NN
(\vct X; \vct Y)_2
& =\vct e^0\wedge\vct X + \mathsf{Y}
\NN
(\vct Y; \beta)_3
& =\vct e^0\wedge\mathsf{Y} + \beta\sigma
,
\end{align}
where
$\sigma=\vct e^1\wedge\vct e^2\wedge\vct e^3$ is the
3-form representing the volume element
and ``$\wedge$'' represents the antisymmetric tensor product.
$\mathsf{Y}=\sum_{j,k=0}^3 \epsilon_{ijk}Y_i\vct e_j\wedge\vct e_k$
is a 2-form (in three dimensional space) derived from $\vct Y$.
$\epsilon_{ijk}$ is the Levi-Civita symbol.
With these, the pair quantities in Table \ref{tab100}
are defined as
\begin{align}
&(\phi, c\vct A)=(-\phi; c\vct A)_1
,
&&(\vct E, c\vct B)=(-\vct E; c\vct B)_2
,\NN  
&(\vct H, c\vct D)=(\vct H; c\vct D)_2
,
&&(\vct J, c\varrho)=(-\vct J; c\varrho)_3
.
\label{eq5010}
\end{align}
The differential operator $\mathsf{d}$ is defined as
\begin{align}
\mathsf{d} = 
\vct{e}^0\wedge\frac{\partial}{\partial x^0}+
\sum_{i=1}^3\vct{e}^i\wedge\frac{\partial}{\partial x^i}
.
\end{align}
and the application to an $n$-form results in
an $(n+1)$-form.
For example, we have
$
\mathsf{d}(\phi, c\vct A)=(\vct E, c\vct B)
$,
which is equivalent to 
$\vct E = -\vct\nabla\phi-\partial\vct A/\partial t$ and
$\vct B = \vct\nabla\times\vct A$.
The successive applications of $\mathsf{d}$
always yield zero ($\mathsf{d}\mathsf{d}=0$), therefore we have
$
\mathsf{d}(\vct E, c\vct B)=0
$, which corresponds to 
$\partial\vct B/\partial t + \vct\nabla\times\vct E=0$,
$\vct\nabla\cdot\vct B=0$.
Furthermore,
$
  \mathsf{d}(\vct H, c\vct D)=(\vct J, c\varrho)
$
yields
$-\partial \vct D/\partial t + \vct\nabla\times\vct H = \vct J$,
$\vct\nabla\cdot\vct D = \varrho$, and
$
\mathsf{d}(\vct J, c\varrho)=0
$
yields the conservation of charge:
$\partial \varrho/\partial t + \vct{\nabla}\cdot\vct J =0$.

Another important notation is
Hodge's star operator ``$*$'', which
converts an $n$-form into a $(4-n)$-form;
\begin{align}
& *(\alpha; \vct X)_1=(\vct X; \alpha)_3,
\NN
& *(\vct X; \vct Y)_2=(-\vct Y; \vct X)_2,
\NN
& *(\vct Y; \beta)_3=(\beta; \vct Y)_1
.
\end{align}
From the second relation and Eq.~(\ref{eq5010}),
the constitutive relations are represented as
$(\vct E, c\vct B)=*Z_0(\vct H, c\vct D)$.

\printtables

\begin{thebibliography}{99}
\bibitem{notation}
Usually, in order to avoid the use of a specific unit
system,
a dimension is represented universally in terms of
$L$ (length), $M$ (mass), $T$ (time), and $I$ (current), 
such as $[f]=LMT^{-2}$.
Here, for simplicity, we use SI units representing dimensions
as $f\simD \U{N}=\U{kg\,m/s^2}$.
\bibitem{dimension}
Consider a rectangular, uniform sheet 
with size $w \times l$.
Two electrodes are attached along the
edges of length $w$.
The resistance $R$ between the electrodes is proportional to
$l$ and inversely proportional to $w$:
$R = \sigma l/w$.
The proportionality constant $\sigma$ is 
called the surface resistivity of the sheet
and $\sigma = Rw/l \simD \U{\Omega m/m}=\U{\Omega}$.
The surface resistivity and the resistance happen to share the
same dimension $\simD\U{\Omega}$ but
their physical meaning is rather different.
Dimensionally the vacuum impedance resembles to the former
because it is introduced in terms of 
the fields of plane wave as
$Z_0=E/H\simD (\U{V/m})/(\U{A/m})=\U{\Omega m/m}$.
See Reference \onlinecite{schelkunoff}.
\bibitem{realcgs}
To be precise the real esu charge $\hat{q}\sub{esu}$ is
\begin{align*}
\hat{q}\sub{esu}=
\frac{\hat{q}}{\sqrt{4\pi}}
\frac{\sqrt{\text{dyn}}\,\U{cm}}{\sqrt{\U{N}}\,\U{m}}
=\frac{\hat{q}}{\sqrt{4\pi\times10^5}}
,
\end{align*}
where $\text{dyn}=\U{g\cdot cm/s^2}$ is the cgs
unit of force.
We will hereafter ignore factors such as $(4\pi\times10^{-5})^{-1/2}$.
\end{thebibliography}
\end{document}